\documentclass[prl,showpacs,twocolumn]{revtex4}%
\usepackage{graphicx}
\usepackage{dcolumn}
\usepackage{bm}
\usepackage{amsmath}
\usepackage{amsfonts}
\usepackage{amssymb}%
\providecommand{\U}[1]{\protect\rule{.1in}{.1in}}
\begin{document}
\title{\textbf{Downfolded Self-Energy of Many-Electron Systems}}
\author{F. Aryasetiawan$^{1,2,3}$, J. M. Tomczak$^{2,3}$, T. Miyake$^{2,3}$, and R.
Sakuma$^{1,3}$}
\affiliation{$^{1}$Graduate School of Advanced Integration Science, Chiba University, 1-33
Yayoi-cho, Inage-ku, Chiba-shi, Chiba, 263-8522 Japan,}
\affiliation{$^{2}$Research Institute for Computational Sciences, AIST, 1-1-1 Umezono,
Tsukuba Central 2, Tsukuba-shi, Ibaraki, 305-8568 Japan,}
\affiliation{$^{3}$Japan Science and Technology Agency, CREST.}

\begin{abstract}
Starting from the full many-body Hamiltonian of interacting electrons the
effective self-energy acting on electrons residing in a subspace of the full
Hilbert space is derived. This subspace may correspond to, for example,
partially filled narrow bands, which often characterize strongly correlated
materials. The formalism delivers naturally the frequency-dependent effective
interaction (the Hubbard $U$) and provides a general framework for
constructing theoretical models based on the Green function language. It also
furnishes a general scheme for first-principles calculations of complex
systems in which the main correlation effects are concentrated on a small
subspace of the full Hilbert space.

\end{abstract}

\maketitle

One of the most important quantities in studying the electronic structure of
materials is the spectral function, a quantity that contains information about
the electronic excitations of the system obtained from photoemission and
inverse photoemission experiment corresponding to the addition and removal of
an electron. A suitable tool to describe the spectral function is the
one-electron Green's function. To calculate the Green function of the full
many-body Hamiltonian is a tremendous task and only recently such
calculations, within the \emph{GW} approximation (GWA) \cite{hedin,ferdi1998},
are feasible thanks to the rapid progress in computer performance. However,
two serious difficulties are hampering further progress: firstly, for many
cases, the system size is too large for realistic \emph{GW} calculations and
secondly, from a theoretical point of view, the GWA may not be sufficient to
treat correlation problems, especially those in materials with strong
correlations containing partially filled narrow bands. These materials with
many intriguing properties have been discovered in recent years \cite{imada}.

The traditional approach for treating such systems is to introduce a model
Hamiltonian, focusing on a small subspace of the full Hilbert space that is
considered to be most relevant for the correlation problem at hand. Notable
examples are the Hubbard model and the Anderson impurity model with the
assumption of a local statically screened Coulomb interaction known as the
Hubbard \emph{U}, which is treated as an adjustable parameter of the models.
These models have given us a lot of physical insight into correlation problems
ranging from the Kondo effect to high temperature superconductivity. The
presence of an adjustable parameter \emph{U}, however, limits the predictive
power of the models, preventing accurate and precise quantitative
calculations. Moreover, these models are not strictly derived from the
many-body Hamiltonian, but rather they are intuitively postulated, which is
not satisfying from theoretical point of view.

The purpose of the present work is to derive from the full many-body
Hamiltonian an exact expression for the self-energy corresponding to a
subspace of the full Hilbert space, expressed in terms of the one-particle
Green's function of the subspace. The motivation for this is that in many
physical problems it is too complicated to treat the full Hilbert space on
equal footing. A sensible and physically motivated approach is to identify a
subspace, where most of the correlations take place. This subspace could be
identified as, for example, partially filled narrow bands crossing the Fermi
level, a typical situation in materials with strong correlations. The task is
then to derive an effective self-energy acting on electrons residing in this
subspace. As a result of the derivation, the effective screened interaction
among electrons belonging to the subspace, i.e., the Hubbard \emph{U}, emerges naturally.

Let us start by defining some basic variables and quantities. The
many-electron Hamiltonian is given by%

\begin{align}
H  &  =\int d^{3}r~\psi^{+}(\mathbf{r})h_{0}(\mathbf{r})\psi(\mathbf{r}%
)\nonumber\\
&  +\frac{1}{2}\int d^{3}rd^{3}r^{\prime}\ \psi^{+}(\mathbf{r})\psi
^{+}(\mathbf{r}^{\prime})v(\mathbf{r-r}^{\prime})\psi(\mathbf{r}^{\prime}%
)\psi(\mathbf{r}), \label{H}%
\end{align}
where $h_{0}$ is the one-particle part of the Hamiltonian and $\psi
(\mathbf{r)}$ is the field operator. We aim at downfolding the many-body
problem onto a subspace of the full Hilbert space, which can consist of, for
example, 3d or 4f orbitals. From now on we refer to the subspace as the $d$
subspace. We first divide the complete field operator into the $d$ field and
the rest, denoted by $d$ and $r$, respectively:%

\begin{equation}
\psi(\mathbf{r})=\psi_{d}(\mathbf{r})+\psi_{r}(\mathbf{r})=\sum_{d}\chi
_{d}(\mathbf{r})c_{d}+\sum_{r}\chi_{r}(\mathbf{r})c_{r}. \label{psi}%
\end{equation}
$\chi_{d}$ and $\chi_{r}$ are the one-particle orbitals and $c_{d}$ and
$c_{r}$ are the associated annihilation operators. We use a convention that
$(\mathbf{r}t)$ is represented by a number. Since we will utilize the
Schwinger functional derivative technique \cite{schwinger} to develop a closed
set of equations for $G^{d}$, where a probing field $\varphi(1)$ is applied to
probe the linear response of the Green function, we define the following
Green's functions in the Dirac or interaction representation%

\begin{equation}
iG^{d}(1,2)=\frac{\left\langle T\left[  S\psi_{d}(1)\psi_{d}^{+}(2)\right]
\right\rangle }{\left\langle S\right\rangle }, \label{Gd}%
\end{equation}

\begin{equation}
iG^{rd}(1,2)=\frac{\left\langle T\left[  S\psi_{r}(1)\psi_{d}^{+}(2)\right]
\right\rangle }{\left\langle S\right\rangle }, \label{Grd}%
\end{equation}

\begin{equation}
S=T\exp\left[  -i\int_{-\infty}^{\infty}dt\ \int d^{3}r\ \varphi
\mathbf{(r}t)\psi^{+}(\mathbf{r}t)\psi(\mathbf{r}t)\right]  .\ \ \ \label{S}%
\end{equation}

Before proceeding further with a detailed derivation, we first summarize
schematically our final result of a closed set of equations for the effective
self-energy and the Green function of the $d$ subspace.%

\begin{align}
\Sigma &  =\Sigma^{d}+\Sigma^{rd}+\Sigma^{drd},\label{Sigmaeff}\\
P^{d}  &  =-iG^{d}\Gamma G^{d},\label{Pd+}\\
\Gamma &  =1+\frac{\delta\Sigma}{\delta G^{d}}G^{d}\Gamma G^{d}%
,\label{vertex0+}\\
W  &  =W^{r}+W^{r}P^{d}W\label{W12+}\\
G^{d}  &  =g^{d}+g^{d}\Sigma G^{d}. \label{GdDyson+}%
\end{align}
In place of the bare Coulomb interaction $v$, we have in (\ref{W12+}) $W^{r}$,
a frequency-dependent effective interaction among electrons living in the
chosen $d$ subspace. This effective interaction is screened by $P^{d}$, the
polarization propagator of the $d$ electrons with a vertex $\Gamma$, yielding
the fully screened interaction $W$ of the full system. $\Gamma$ is an
effective vertex acting on the $d$ subspace and consequently different from
the full vertex. If the subspace is isolated and identified as the Hilbert
space of a Hubbard model and $W^{r}$ is approximated by a local and static
interaction (the Hubbard $U$), $\Sigma^{d}$ is then the exact self-energy of
the Hubbard model. The explicit expressions for the self-energies will be
shown later. In addition to $\Sigma^{d}$ the true effective self-energy
$\Sigma$ contains a term, $\Sigma^{rd}$, which is the self-energy arising from
the $r$ subspace acting on the $d$ subspace and a hopping term $\Sigma^{drd}$
representing hybridization between the $d$ and $r$ subspaces and scattering
processes of an electron from the $d$ to the $r$ and back to the $d$ subspace.
Thus, $\Sigma$ is not merely a projection of the full self-energy onto the $d$
subspace. As will be explained later, for a \emph{given} $W^{r}$ the effective
self-energy is a functional of $G^{d}$ only. The last equation is the Dyson
equation with a noninteracting Green's function $g^{d}$ of the $d$ subspace.

We now proceed with the derivation of the above set of equations and use the
convention that repeated indices are summed and repeated variables are
integrated, unless they appear on both sides of the equation. From the
Heisenberg equation of motion for the field operators $\psi_{d}$ and $\psi
_{r}$ we obtain the equations of motion for $G^{d}$ and $G^{rd}$:%

\begin{align}
&  i\frac{\partial}{\partial t_{1}}G^{d}(1,2)-\Delta_{d}(1,3)h^{0}%
(3)[G^{d}(3,2)+G^{rd}(3,2)]\nonumber\\
&  -\Delta_{d}(1,3)\left[  M^{d}(3,4)+M^{rd}(3,4)\right]  G^{d}(4,2)=\Delta
_{d}(1,2), \label{dtGd}%
\end{align}

\begin{align}
&  i\frac{\partial}{\partial t_{1}}G^{rd}(1,2)-\Delta_{r}(1,3)h^{0}%
(3)[G^{d}(3,2)+G^{rd}(3,2)]\nonumber\\
&  -\Delta_{r}(1,3)\left[  M^{d}(3,4)+M^{rd}(3,4)\right]  G^{d}(4,2)=0,
\label{dtGrd}%
\end{align}
where%

\begin{equation}
\Delta_{i}(1,2)=\chi_{i}(\mathbf{r}_{1})\chi_{i}^{\ast}(\mathbf{r}_{2}%
)\delta(t_{1}-t_{2}),\ \ \ (i=d,r), \label{Dd}%
\end{equation}
and we have \emph{defined} the mass operators according to%

\begin{align}
&  M^{d}(1,4)G^{d}(4,2)\nonumber\\
&  =-iv(1-3)~\left\langle T\left[  \psi^{+}(3)\psi(3)\psi_{d}(1)\psi_{d}%
^{+}(2)\right]  \right\rangle , \label{Md}%
\end{align}

\begin{align}
&  M^{rd}(1,4)G^{d}(4,2)\nonumber\\
&  =-iv(1-3)~\left\langle T\left[  \psi^{+}(3)\psi(3)\psi_{r}(1)\psi_{d}%
^{+}(2)\right]  \right\rangle . \label{Mrd}%
\end{align}
It is important to realize that the mass operator $M^{rd}$ is defined with
respect to $G^{d}$, rather than $G^{rd}$. Noting that $\delta S/\delta
\varphi(3)=-iT\left[  S\psi^{+}(3)\psi(3)\right]  $ the four-field term can be
related to $\delta G^{d}/\delta\varphi$ and $\delta G^{rd}/\delta\varphi$
\cite{schwinger,hedin,ferdi1998}. Thus, defining the self-energy as the mass
operator less the Hartree potential $V_{H}$ we obtain%

\begin{equation}
\Sigma^{j}(1,4)G^{d}(4,2)=iW(3,1)\frac{\delta G^{j}(1,2)}{\delta
V(3)},\ \ (j=d,rd) \label{Md1}%
\end{equation}
We have defined $V=V_{H}+\varphi$ and the screened Coulomb interaction
$W=(\delta V/\delta\varphi)v=\varepsilon^{-1}v.$

To derive an effective self-energy operator for $G^{d}$, we need to eliminate
$G^{rd}$. At this stage it is convenient to work in the basis representation.
We denote a basis in the $d$ subspace by Greek letters ($\alpha$) and in the
rest of the subspace by small roman letters ($n$) whereas capital letters
($N$) denote the full Hilbert space. Note also that $G^{d}$ has only matrix
elements in the $d$ subspace whereas $G^{rd}$ has only off-diagonal elements
between the $d$-$r$ subspaces. However, the self-energies $\Sigma^{d}$ and
$\Sigma^{rd}$ can have matrix elements between the $d$-$d$ and $d$-$r$ subspaces.

The equations of motion of $G^{d}$ and $G^{rd}$ in frequency space become,%

\begin{equation}
\lbrack\omega\delta_{\alpha\gamma}-h_{\alpha\gamma}-\Sigma_{\alpha\gamma}%
^{d}(\omega)-\Sigma_{\alpha\gamma}^{rd}(\omega)]G_{\gamma\beta}^{d}%
(\omega)-h_{\alpha n}G_{n\beta}^{rd}(\omega)=\delta_{\alpha\beta},
\label{G(w)}%
\end{equation}

\begin{equation}
\lbrack\omega\delta_{nm}-h_{nm}]G_{m\beta}^{rd}(\omega)-[h_{n\gamma}%
+\Sigma_{n\gamma}^{d}(\omega)+\Sigma_{n\gamma}^{rd}(\omega)]G_{\gamma\beta
}^{d}(\omega)=0, \label{g(w)}%
\end{equation}
where the Hartree potential has been absorbed into $h=h^{0}+V_{H}$. Note that
$h$ is static unless a probing time-dependent field $\varphi(\mathbf{r}t)$ is
applied. Using the second equation we eliminate $G^{rd}$:%

\begin{equation}
G_{n\beta}^{rd}(\omega)=g_{nm}^{r}(\omega)[h_{m\gamma}+\Sigma_{m\gamma}%
^{d}(\omega)+\Sigma_{m\gamma}^{rd}(\omega)]G_{\gamma\beta}^{d}(\omega),
\label{Grdw}%
\end{equation}
where $g_{nm}^{r}=[\omega-h]_{nm}^{-1}$, which is a noninteracting Green's
function of the $r$ subspace. A small letter $g$ is used to denote a
noninteracting Green's function. The effective self-energy for $G^{d}$ is
defined as follows.%

\begin{equation}
\lbrack\omega\delta_{\alpha\gamma}-h_{\alpha\gamma}-\Sigma_{\alpha\gamma
}(\omega)]G_{\gamma\beta}^{d}(\omega)=\delta_{\alpha\beta}, \label{Dysonw}%
\end{equation}

\begin{equation}
\Sigma_{\alpha\gamma}(\omega)=\Sigma_{\alpha\gamma}^{d}(\omega)+\Sigma
_{\alpha\gamma}^{rd}(\omega)+\Sigma_{\alpha\gamma}^{drd}(\omega),
\label{Sigmaw}%
\end{equation}

\begin{equation}
\Sigma_{\alpha\gamma}^{drd}(\omega)=h_{\alpha n}g_{nm}^{r}(\omega)[h_{m\gamma
}+\Sigma_{m\gamma}^{d}(\omega)+\Sigma_{m\gamma}^{rd}(\omega)].
\label{Sigmadrd}%
\end{equation}

We are now in the position to calculate $\Sigma^{d}$ and $\Sigma^{rd}.$ We
first note that $G^{d}(G^{d})^{-1}=\Delta_{d}$ since $G^{d}$ is only defined
within the $d$ subspace and $\delta V_{MN}(t)/\delta V(\mathbf{r}^{\prime
}t^{\prime})=\chi_{M}^{\ast}(\mathbf{r}^{\prime}\mathbf{)}\chi_{N}%
(\mathbf{r}^{\prime}\mathbf{)}\delta(t-t^{\prime})$. Multiplying both sides of
(\ref{Md1}) on the right by $(G^{d})^{-1}$, and using the identity
$G^{d}\delta(G^{d})^{-1}+\delta G^{d}(G^{d})^{-1}=0$ we obtain%

\begin{equation}
\Sigma_{N\beta}^{d}(t_{1}-t_{2})=iW_{N\eta,\mu\nu}(t_{3}-t_{1})G_{\eta\kappa
}^{d}(t_{1}-t_{4})\Gamma_{\mu\nu}^{\kappa\beta}(t_{4}-t_{2};t_{3}).
\label{Sigmadt}%
\end{equation}
The vertex $\Gamma$ is given by%

\begin{align}
&  \Gamma_{\mu\nu}^{\alpha\beta}(t_{1}-t_{2},t_{3})\nonumber\\
&  =-\frac{\delta(G^{d})_{\alpha\beta}^{-1}(t_{1}-t_{2})}{\delta V_{\mu\nu
}(t_{3})}\nonumber\\
&  =\delta_{\alpha\mu}\delta_{\beta\nu}\delta(t_{1}-t_{2})\delta(t_{1}%
-t_{3})\nonumber\\
&  \ \ \ \ +\frac{\delta\Sigma_{\alpha\beta}(t_{1}-t_{2})}{\delta
G_{\gamma\eta}^{d}(t_{4})}G_{\gamma\kappa}^{d}(t_{4}-t_{5})\Gamma_{\mu\nu
}^{\kappa\lambda}(t_{5}-t_{6},t_{3})G_{\lambda\eta}^{d}(t_{6}).
\label{vertexeq}%
\end{align}
The last line has been obtained by using a chain rule $\delta\Sigma/\delta
V=(\delta\Sigma^{d}/\delta G^{d})(\delta G^{d}/\delta V)$ and the identity
$\delta G^{d}=-G^{d}\delta(G^{d})^{-1}G^{d}$.

Similarly%

\begin{align}
\Sigma_{K\beta}^{rd}(t_{1}-t_{2})  &  =iW_{Km,MN}(t_{3}-t_{1})\nonumber\\
&  \times\frac{\delta G_{m\gamma}^{rd}(t_{1}-t_{4})}{\delta V_{MN}(t_{3}%
)}(G^{d})_{\gamma\beta}^{-1}(t_{4}-t_{2}). \label{Sigmard}%
\end{align}
To calculate $\Sigma^{rd}$ we need to take the functional derivative of
$G_{n\alpha}^{rd}$ with respect to the probing field. However, $G_{n\alpha
}^{rd}$ cannot be inverted and consequently the usual trick of using the
identity $\delta(GG^{-1})=0$ does not apply. We therefore must calculate
$\delta G^{rd}/\delta V$ directly. To do this we use $G^{rd}(\omega)$ in
(\ref{Grdw}), Fourier transform back into time domain and perform the
functional derivative. Calculating $\delta G^{rd}/\delta V$ and substituting
the result in (\ref{Sigmard}) yields%
\begin{align}
&  \Sigma_{N\beta}^{rd}(t_{1}-t_{2})\nonumber\\
&  =\ \Sigma_{Nl}^{GW(r)}(t_{1}-t_{3})g_{ln}^{r}(t_{3}-t_{5})\nonumber\\
&  \times\left[  h_{n\beta}\delta(t_{5}-t_{2})+\Sigma_{n\beta}^{d}(t_{5}%
-t_{2})+\Sigma_{n\beta}^{rd}(t_{5}-t_{2})\right] \nonumber\\
\ \ \ \ \ \  &  +\Sigma_{N\beta}^{GW(r)}(t_{1}-t_{2})\nonumber\\
&  +iW_{Nm,\mu\nu}(t_{3}-t_{1})g_{mn}^{r}(t_{1}-t_{5})\nonumber\\
&  \ \ \ \ \ \times\left[  \frac{\delta\Sigma_{n\beta}^{d}(t_{5}-t_{2}%
)}{\delta V_{\mu\nu}(t_{3})}+\frac{\delta\Sigma_{n\beta}^{rd}(t_{5}-t_{2}%
)}{\delta V_{\mu\nu}(t_{3})}\right] \nonumber\\
&  +iW_{Nm,\mu\nu}(t_{3}-t_{1})g_{mn}^{r}(t_{1}-t_{5})\nonumber\\
&  \times\left[  h_{n\lambda}\delta(t_{5}-t_{6})+\Sigma_{n\lambda}^{d}%
(t_{5}-t_{6})+\Sigma_{n\lambda}^{rd}(t_{5}-t_{6})\right] \nonumber\\
&  \ \ \ \ \ \ \ \ \ \ \ \ \ \ \ \ \ \ \ \ \times G_{\lambda\gamma}^{d}%
(t_{6}-t_{4})\Gamma_{\mu\nu}^{\gamma\beta}(t_{4}-t_{2};t_{3}),
\label{Sigmardt}%
\end{align}
where%

\begin{equation}
\Sigma_{MN}^{GW(r)}(t)=ig_{mn}^{r}(t)W_{Mm,nN}(t). \label{SigmaGWr}%
\end{equation}
Finally, the polarization propagator of the $d$ subspace is given by%

\begin{align}
P^{d}(1,2)  &  =\frac{\delta\rho^{d}(1)}{\delta V(2)}=-i\frac{\delta
G^{d}(1,1^{+})}{\delta V(2)}\nonumber\\
&  =\chi_{\alpha}(\mathbf{r}_{1})\chi_{\beta}^{\ast}(\mathbf{r}_{1}%
)P_{\alpha\beta,\mu\nu}^{d}(t_{1}-t_{2})\chi_{\mu}(\mathbf{r}_{2})\chi_{\nu
}^{\ast}(\mathbf{r}_{2}), \label{Pd0}%
\end{align}

\begin{equation}
P_{\alpha\beta,\mu\nu}^{d}(t_{1}-t_{2})=-iG_{\alpha\gamma}^{d}(t_{1}%
-t_{3})\Gamma_{\mu\nu}^{\gamma\eta}(t_{3}-t_{4},t_{2})G_{\eta\beta}^{d}%
(t_{4}-t_{1}^{+}). \label{Pdbas}%
\end{equation}
Writing the full polarization of the system as a sum of the $d$- and
$r$-subspace polarization, $P=P^{d}+P^{r},$ it is straightforward to verify
that with%

\begin{equation}
W^{r}=v+vP^{r}W^{r}=[1-vP^{r}]^{-1}v, \label{Wr}%
\end{equation}
Eq. (\ref{W12+}) is equivalent to $W=v+vPW$. This completes our derivation of
the closed set of equations for $G^{d}$ and the effective self-energy, which
is schematically given in Eqs. (\ref{Sigmaeff}), (\ref{Pd+}), (\ref{vertex0+}%
), (\ref{W12+}), and (\ref{GdDyson+}), and in detail in Eqs. (\ref{Sigmaw}),
(\ref{Pdbas}), (\ref{vertexeq}), (\ref{W12+}), and (\ref{Dysonw}). This set of
equations may be regarded as a set of downfolded Hedin's equations for a
\emph{given} $W^{r}$. This quantity may be calculated, for example, within the
constrained random-phase approximation (cRPA) scheme \cite{ferdi2004}. For a
given screened interaction $W^{r}$, the self-energies and hence the effective
self-energy are functionals of $G^{d}$ only, as can be seen in Eqs.
(\ref{Sigmadt}), (\ref{Sigmardt}), and (\ref{Sigmadrd}). The dependence on
$G^{rd}$ has been eliminated and there is no explicit dependence on $G^{r}$.
It is important to note that we have defined the effective self-energy such
that%
\begin{equation}
\Sigma_{\alpha\gamma}G_{\gamma\beta}^{d}=\Sigma_{\alpha\gamma}^{F}%
\mathcal{G}_{\gamma\beta}+\Sigma_{\alpha n}^{F}\mathcal{G}_{n\beta}
\label{Sigmaequiv}%
\end{equation}
where $\Sigma^{F}$ and $\mathcal{G}$ are respectively the self-energy and the
Green function of the \emph{full} Hilbert space. The first term on the
right-hand-side is confined to the $d$ subspace and the second represents a
coupling between the $d$ and $r$ subspaces. Thus, the self-energy
$\Sigma_{\alpha\beta}$ is not a simple projection of the full self-energy onto
the $d$ subspace but it contains the effects of the off-diagonal matrix
elements of $\Sigma^{F}$ and $\mathcal{G}$ between the $d$ and $r$ subspaces.
However, $G^{d}$ is equivalent to the projection of $\mathcal{G}$ onto the $d$
subspace as can be seen in (\ref{Gd}).

It is instructive to compute the effective self-energy in our formalism within
the GWA,\emph{ }where the vertex corrections $\delta\Sigma^{d}/\delta V$ and
$\delta\Sigma^{rd}/\delta V$ are neglected.%

\begin{align}
\Sigma_{\alpha\beta}^{GW}(\omega)  &  =\Sigma_{\alpha\beta}^{GW(d)}%
(\omega)+\Sigma_{\alpha\beta}^{GW(r)}(\omega)\nonumber\\
&  +\Sigma_{\alpha l}^{GW(r)}(\omega)g_{ln}^{r}(\omega)[h_{n\beta}%
+\Sigma_{n\beta}^{d}(\omega)+\Sigma_{n\beta}^{rd}(\omega)]\nonumber\\
&  +g_{mn}^{r}(\omega)\left[  h_{n\lambda}+\Sigma_{n\lambda}^{d}%
(\omega)+\Sigma_{n\lambda}^{rd}(\omega)\right] \nonumber\\
&  \ \ \ \ \ \times G_{\lambda\gamma}^{d}(\omega)W_{\alpha m,\gamma\beta
}(\omega)\nonumber\\
&  +\Sigma_{\alpha\beta}^{drd}(\omega), \label{SigmaGW}%
\end{align}
where $\Sigma_{\alpha\beta}^{GW(r)}$ and $\Sigma_{\alpha\beta}^{drd}$ are
given in (\ref{SigmaGWr}) and (\ref{Sigmadrd}), and%

\begin{equation}
\Sigma_{\alpha\beta}^{GW(d)}(\omega)=(i/2\pi)G_{\eta\kappa}^{d}(\omega
+\omega^{\prime})W_{\alpha\eta,\kappa\beta}(\omega^{\prime}). \label{SigmaGWd}%
\end{equation}
The first two terms correspond to the projection of the full GW self-energy
onto the $d$ subspace. As discussed in the previous paragraph, our effective
self-energy contains the effects of the off-diagonal self-energy between the
$d$ and $r$ subspaces: The third and fourth terms with the square brackets are
the self-energy contribution from the $r$ subspace on the $d$ subspace. The
term $g_{mn}^{r}(\omega)\left[  h_{n\lambda}+\Sigma_{n\lambda}^{d}%
(\omega)+\Sigma_{n\beta}^{rd}(\omega)\right]  G_{\lambda\gamma}^{d}(\omega)$
may be interpreted as a self-energy correction to $G^{d}$ arising from the
off-diagonal elements of $h$, $\Sigma^{d}$, and $\Sigma^{rd}$. This together
with $W_{\alpha m,\gamma\beta}(\omega)$ forms a \emph{GW }self-energy
correction. The last term $\Sigma^{drd}$ is the hybridization term
representing hoppings of electrons from the $d$ to the $r$ subspace and back
to the $d$ subspace. Using the \emph{GW} self-energy and keeping \emph{W}
fixed it becomes feasible to solve the vertex equation in (\ref{vertexeq})
since the subspace of interest is usually much smaller than the full Hilbert
space. The resulting vertex $\Gamma$ can then be inserted into the self-energy
in (\ref{Sigmadt}) to obtain an improved self-energy beyond the GWA.

The set of equations for the downfolded self-energy provides a general
framework for constructing theoretical models in the Green function language.
It offers an alternative to model Hamiltonians, which cannot readily take into
account frequency-dependent interactions. Rather than first mapping the full
Hamiltonian to a model Hamiltonian with a static $U$ and then solving the
model using the Green function technique, the present formalism allows for a
direct route to the self-energy with a full frequency-dependent effective
interaction. When $W^{r}$ is approximated by a local and static value it
becomes evident that the set of equations may be interpreted as equivalent to
the Hubbard model when $\Sigma^{rd}$ and $\Sigma^{drd}$ are neglected.
Consequently, these two self-energy terms have no counterparts in the Hubbard
model. Two ingredients are therefore missing in the Hubbard model, namely, the
frequency dependence of $U$, which can be important \cite{ferdi2004}, and the
self-energy effects arising from the $r$ subspace. In the Anderson impurity
model, the orbitals of the impurity may be viewed as forming the $d$ subspace
and the influence of the $r$ subspace is included as a hybridization term,
which may be identified as the first term in (\ref{Sigmadrd}). However, the
model does not include the self-energy effects $\Sigma^{rd}$ arising from the
$r$ subspace and as in the Hubbard model, the effective interaction is local
and static.

Finally let us consider some possible general applications. The present
formalism can be used to improve the Hubbard or the Anderson impurity model by
providing better parameters and by including the missing self-energy terms and
the effects of the frequency-dependent $W^{r}$ within, for example, the GWA.
The formalism also provides a route for numerical simplification of the GWA,
by focusing on a certain subspace and treating the rest in an approximate,
e.g., mean-field scheme. This is highly desirable since applications of the
GWA to complex systems are computationally demanding. Also, the formalism may
facilitate a tractable way of going beyond the GWA fully from first-principles
by inclusion of vertex corrections on the given subspace only.

In conclusion, we have derived from the full many-body Hamiltonian a closed
set of equations for the effective self-energy acting on a subspace of the
full Hilbert space. The effective frequency-dependent interaction or the
Hubbard $U$ appears naturally in this formalism. Since the equations are
exact, they provide a general framework for handling complex systems in which
the main correlation effects are concentrated on a certain subspace, such is
the case in many correlated materials characterized by partially filled narrow
bands. It offers an alternative to model Hamiltonians by constructing
theoretical models in the Green function language allowing for a direct access
to the self-energy and a natural inclusion of frequency-dependent
interactions, usually not accounted for in conventional model Hamiltonians.

We thank Silke Biermann for many fruitful discussions and suggestions.

\end{document}